\documentclass[final]{svjour2}
\usepackage{bm}
\newcommand{\kB}{k_{\rm B}}
\newcommand{\partialr}{\frac{\partial}{\partial\bm{r}}}
\newcommand{\aver}[1]{\left\langle#1\right\rangle}
\newcommand{\averr}[1]{\left\langle\!\left\langle #1 \right\rangle\!\right\rangle}

\journalname{Journal of Statistical Physics}

\begin{document}
\bibliographystyle{spmpsci}

\title{Nonequilibrium Thermodynamics of the First and Second Kind: Averages and Fluctuations}
\titlerunning{Nonequilibrium Thermodynamics}

\author{Hans Christian \"Ottinger}
\authorrunning{H.\,C.~\"Ottinger}

\institute{H.\,C.~\"Ottinger \at ETH Z\"urich, Department of Materials, Polymer Physics, HCI H 543, CH-8093 Z\"urich, Switzerland\\\email{hco@mat.ethz.ch}}

\date{\today}

\maketitle

\begin{abstract}
We compare two approaches to nonequilibrium thermodynamics, the two-generator bracket formulation of time-evolution equations for averages and the macroscopic fluctuation theory, for an isothermal driven diffusive system under steady state conditions. The fluctuation dissipation relations of both approaches play an important role for a detailed comparison. The nonequilibrium Helmholtz free energies introduced in these two approaches differ as a result of boundary conditions. A Fokker-Planck equation derived by projection operator techniques properly reproduces long range fluctuations in nonequilibrium steady states and offers the most promising possibility to describe the physically relevant fluctuations around macroscopic averages for time-dependent nonequilibrium systems.
\end{abstract}




\section{Introduction}
There appears to be a widespread believe that time-dependent far-from-equilibrium systems are too complicated to be dealt with and that one should hence develop a thermodynamic framework for nonequilibrium steady states first. With a strong focus on phenomenological thermodynamics, we just mention the fundamental attempts to establish a general theory of nonequilibrium steady states made by Oono and Paniconi (1998) \cite{OonoPaniconi98}, Sasa and Tasaki (2006) \cite{SasaTasaki06}, Bertini \emph{et al.} (2002--2009) \cite{Bertinietal02,Bertinietal04,Bertinietal06,Bertinietal09}, and Taniguchi and Cohen (2007) \cite{TaniguchiCohen07}. This list of efforts is by no means complete, but it gives a flavor of the variety of different approaches; many further references can be found in the above papers. Most of these groups look for support for their phenomenological thermodynamic approaches from statistical mechanics. Among the important insights into the statistical mechanics of nonequilibrium steady states, we mention the work of Derrida \emph{et al.} \cite{Derridaetal01,Derridaetal02} on the fluctuations in an exactly solvable model of a driven diffusive system, fluctuation theorems \cite{Evansetal93,GallavottiCohen95,Gallavotti98}, and the fundamental concept of SRB measures (Sinai, Ruelle, Bowen \cite{Sinai72,Bowen,BowenRuelle75,Ruelle76}) defined on the attractors of chaotic systems (see also the reviews \cite{EckmannRuelle85,Ruelle99,Young02}).

While the relative simplicity of steady state systems seems to be indisputable, there may also be some reasons to believe that looking in a sufficiently abstract way at the general problem of nonequilibrium time evolution might help to clarify the structure of nonequilibrium thermodynamics. First of all, the formulation of time evolution can be given in terms of illuminating geometric structures for generating trajectories. Second, in the evolution equations one can easily separate reversible and irreversible contributions, where the hallmark of reversibility is the possibility of a Hamiltonian formulation. Third, the fundamental thermodynamic concepts of energy and entropy reappear naturally in the leading parts of nonequilibrium dynamics as generators of reversible and irreversible evolution, respectively. Fourth, in time-dependent local field theories one can easily separate bulk from boundary effects, which is difficult, if not impossible, in a steady state system because the influence of the boundary conditions penetrates through the entire system. Fifth, and perhaps most importantly, we have the machinery of the projection operator formalism \cite{Zwanzig61,Mori65,Mori65a,Robertson66,Grabert}, which relies on a clear separation of time scales. Separation of time scales is a trump card gamed away in the theory of steady state systems. Detailed elaborations on all these issues can be found in the textbook \cite{hcobet} which summarizes a stream of developments that started with three pioneering letters \cite{Kaufman84,Morrison84,Grmela84} in 1984, was concretized into the clarifying textbook \cite{BerisEdwards}, and formulated as a general framework in the papers \cite{hco99,hco100,hco101}.

This paper is organized as follows. We first summarize the two-generator bracket formulation of the general evolution equations for nonequilibrium systems (Sec.~\ref{secGENERIC}) and we then illustrate the ideas of this framework by deriving the equations for a driven diffusive system (Sec.~\ref{secdiffsys}). In Sec.~\ref{secMFT}, macroscopic fluctuation theory is presented in the context of the driven diffusive system. The different approaches to nonequilibrium thermodynamics are compared in Sec.~\ref{secavefluc}, which leads to the distinction between thermodynamics of the first and second kind based on averages and fluctuations, respectively. In Sec.~\ref{secFDR}, fluctuation dissipation relations are used to establish deeper connections between the two-generator bracket formulation and the macroscopic fluctuation theory. We give an explicit example of a calculation of long range correlations in a nonequilibrium steady state within the phenomenological two-generator bracket approach. A summary and discussion conclude the paper (Sec.~\ref{secsumdisc}).

\section{GENERIC}\label{secGENERIC}
Time-evolution equations for nonequilibrium systems have a well-defined structure in which reversible and irreversible contributions can be specified separately. In particular, the reversible contribution is generally assumed to be of the Hamiltonian form and hence requires an underlying geometric structure (a Poisson bracket) which reflects the idea that the reversible time evolution should be ``under mechanistic control.'' The remaining irreversible contribution is generated by the nonequilibrium entropy by means of a dissipative bracket. The nonequilibrium energy and entropy landscapes are introduced through their roles as generators of reversible and irreversible dynamics in the space of nonequilibrium variables; they are associated with the evolution of averages. It has been shown in Sec.~II.B.5 of \cite{hco99} how equilibrium thermodynamics in its familiar form arises from these generators of time evolution.

Our discussion is based on the general equation for the nonequilibrium reversible-irreversible coupling (GENERIC) for the time-evolution of nonequilibrium systems \cite{hco99,hco100,hcobet}. If $A$ is an arbitrary observable, that is, a sufficiently regular real-valued function or functional of a set of variables $x$ required for a complete description of a given nonequilibrium system, the time evolution of $A$ is given by
\begin{equation} \label{brackform}
  \frac{dA}{dt} = \{A,E\} + [A,S] .
\end{equation}
The observables $E$ and $S$ generating time evolution are the total energy and entropy, and $\{\cdot,\cdot\}$ and $[\cdot,\cdot]$ are Poisson and dissipative brackets, respectively. The bracket of two observables $A$ and $B$ is another observable with a linear dependence on $A$ and $B$ (a more complete characterization of Poisson and dissipative brackets is given in Eqs.~(\ref{condLasym})--(\ref{condLJacobi}) below). The two contributions to the time evolution of $A$ generated by the total energy $E$ and the entropy $S$ in Eq.~(\ref{brackform}) are referred to as the reversible and irreversible contributions, respectively. Equation (\ref{brackform}) is supplemented by the complementary degeneracy requirements
\begin{equation} \label{LSconsistency}
  \{S,A\} = 0 ,
\end{equation}
and
\begin{equation} \label{MEconsistency}
  [E,A] = 0 ,
\end{equation}
which hold for all observables $A$. The requirement that the entropy is a degenerate functional of the Poisson bracket expresses the reversible nature of the first contribution to the dynamics: the functional form of the entropy is such that it cannot be affected by the Poisson bracket contribution to the dynamics, no matter which observable $A$ is used as a generator of reversible dynamics. The existence of degenerate observables is a hallmark of coarse graining because the Poisson bracket associated with the symplectic structure of atomistic equations is non-degenerate and hence does not allow for the existence of an entropy on the purely reversible atomistic level. The requirement that the energy is a degenerate functional of the dissipative bracket expresses the conservation of the total energy by the dissipative contribution to the dynamics in a closed system.

For completeness, we give the defining properties of Poisson and dissipative brackets. The Poisson bracket possesses the antisymmetry property
\begin{equation} \label{condLasym}
  \{A,B\}=-\{B,A\} ,
\end{equation}
and satisfies the product or Leibniz rule
\begin{equation} \label{Leibnizrule}
  \{A B,C\} = A \{B,C\} + B \{A,C\} ,
\end{equation}
as well as the Jacobi identity
\begin{equation} \label{condLJacobi}
  \{A,\{B,C\}\}+\{B,\{C,A\}\}+\{C,\{A,B\}\}=0 ,
\end{equation}
where $A$, $B$, and $C$ are arbitrary observables. These properties are well-known from the Poisson brackets of classical mechanics, and they express the essence of reversible dynamics. The Jacobi identity (\ref{condLJacobi}), which is a highly restrictive condition for formulating proper reversible dynamics, expresses the invariance of Poisson brackets in the course of time (time-structure invariance).

The dissipative bracket satisfies the symmetry condition (for a more sophisticated discussion of the Onsager-Casimir symmetry properties of the dissipative bracket, see Sections 3.2.1 and 7.2.4 of \cite{hcobet})
\begin{equation} \label{condMsym}
  [A,B]=[B,A] ,
\end{equation}
and the non-negativeness condition
\begin{equation} \label{condMpos}
  [A,A] \ge 0 .
\end{equation}
This non-negativeness condition, together with the degeneracy requirement (\ref{LSconsistency}), guarantees that the entropy is a nondecreasing function of time,
\begin{equation} \label{increntrop}
  \frac{dS}{dt} = [S,S] \ge 0 .
\end{equation}
The condition (\ref{condMpos}) may hence be regarded as a strong formulation of the second law of nonequilibrium thermodynamics.

In practical calculations, it is often convenient to formulate GENERIC in terms of Poisson and friction operators instead of brackets \cite{hco100,hcobet}. This is the analogue of formulating Hamilton's equations of motion in terms of a symplectic matrix rather than in terms of Poisson brackets. More precisely, one generally writes
\begin{equation} \label{Pbrackrepr}
  \{A,B\} = \frac{\delta A}{\delta x} L \frac{\delta B}{\delta x} ,
\end{equation}
and
\begin{equation} \label{dbrackrepr}
  [A,B] = \frac{\delta A}{\delta x} M \frac{\delta B}{\delta x} ,
\end{equation}
where $L$ is the Poisson operator and $M$ is the friction operator. In case that $L$ and $M$ are differential operators, boundary terms need to be discussed separately, as illustrated in the subsequent section on diffusive systems. The Leibniz rule (\ref{Leibnizrule}) then follows automatically. The time-evolution equations for the system variables $x$ implied by Eq.~(\ref{brackform}) can be expressed in the form
\begin{equation}\label{GENERICLM}
  \frac{d x}{d t} = L \frac{\delta E}{\delta x}
  + M \frac{\delta S}{\delta x} .
\end{equation}
If there is a constant $T_0$ with dimensions of temperature, then we can introduce the Helmholtz free energy,
\begin{equation}\label{FGENdef}
    F = E - T_0 S ,
\end{equation}
and rewrite Eq.~(\ref{GENERICLM}) in terms of the single generator $F$,
\begin{equation}\label{GENERICLM1}
  \frac{d x}{d t} = L \frac{\delta F}{\delta x}
  - \frac{1}{T_0} M \frac{\delta F}{\delta x} ,
\end{equation}
where the mutual degeneracy requirements (\ref{LSconsistency}) and (\ref{MEconsistency}) have been used. Double generator ($E$, $S$) and single generator ($F$) formulations of nonequilibrium thermodynamics have been compared in great detail \cite{Edwards98,hco110,Beris01}.

The safest case of a constant temperature arises in quantum gravity where $T_0$ is the Planck temperature, that is, a constant of nature \cite{hco180}. In the present context, however, the idea is to consider isothermal systems. Although very common, the assumption of isothermal conditions is subtle and may even obscure fundamental discussions of nonequilibrium systems. As entropy production is a key feature of nonequilibrium systems, it would be preferable to include heat conduction into the discussion. Steady states are naturally associated with nonuniform temperature profiles and the transport of entropy through the boundaries of an open system. The treatment of the required boundary conditions within the GENERIC framework, which is crucial for the discussion of driven systems, has been discussed in a number of recent publications \cite{hco162,hco171,hco172,hco188}. In the context of SRB measures, isothermal conditions are achieved by including a thermostat into the equations of motion, which might be considered as undesirable for an attempt to clarify the conceptual foundations of nonequilibrium thermodynamics.

The structure of GENERIC can be obtained by the projection operator method \cite{hco101,hco131,hcobet}. This technique relies on nonequilibrium statistical ensembles, in particular, generalized microcanonical and canonical ensembles, and projectors on the spaces of slow and fast variables. As a result, one obtains practical recipes for calculating the GENERIC building blocks by means of statistical mechanics \cite{hco173}. In this paper, however, we focus on the phenomenological approach to nonequilibrium thermodynamics.

\section{Driven Diffusive System}\label{secdiffsys}
To illustrate the GENERIC framework, we derive the evolution equation for a diffusive system. As our only system variable $x$, we choose the mass density field $x = \rho(\bm{r})$ of the diffusing species on a domain $\Lambda$. We further assume isothermal conditions at temperature $T_0$. For this assumption to be meaningful, we consider an athermal hard-sphere system for which the internal energy density is entirely of kinetic origin and hence of the ideal gas form,
\begin{equation}\label{intenergydens}
    \epsilon(\rho) = \frac{3}{2} \frac{\kB T_0}{m} \rho ,
\end{equation}
where $\kB$ is Boltzmann's constant and $m$ is the mass of a spherical particle. An important observation is that the derivative of $\epsilon(\rho)$ with respect to $\rho$ is constant. The entropy density of a hard-sphere system is given by
\begin{equation}\label{entropydens}
    s(\rho) = \frac{\kB}{m} \rho \ln \left[ T_0^{3/2} \frac{R_0(\rho)}{\rho} \right] ,
\end{equation}
where $R_0(\rho)$ is a given function of $\rho$. For an ideal gas, that is, in the limit of vanishing particle radius, $R_0(\rho)$ is a constant with proper dimensions (to be constructed with the help of Planck's constant). The derivative of $s(\rho)$ with respect to $\rho$ at constant $T_0$ is associated with the chemical potential $\mu$ per unit mass according to
\begin{equation}\label{chempotential}
    \frac{\mu}{T_0} = \frac{1}{T_0} \frac{\partial f}{\partial \rho}
    = \frac{1}{T_0} \frac{\partial \epsilon}{\partial \rho}
    - \frac{\partial s}{\partial \rho}
    = - \frac{\partial s}{\partial \rho} + \frac{3}{2} \frac{\kB}{m} ,
\end{equation}
where the Helmholtz free energy density for our isothermal system is given by
\begin{equation}\label{freeenergy}
    f(\rho) = \epsilon(\rho) - T_0 s(\rho) .
\end{equation}

The total energy and entropy are obtained as the integrals of their densities over the domain $\Lambda$,
\begin{equation}\label{ESexpr}
    E = \int_\Lambda \epsilon(\rho(\bm{r})) d^3 r , \qquad
    S = \int_\Lambda s(\rho(\bm{r})) d^3 r ,
\end{equation}
and, for these simple functionals, the functional derivatives of $E$ and $S$ with respect to $\rho(\bm{r})$ are simply given by the partial derivatives of the respective densities.

As a next step, we need to introduce the Poisson and friction operators $L(x)$ and $M(x)$. For hydrodynamic systems, the only reversible effect is convection. As we consider a purely diffusive system and we have eliminated the velocity field from the description, the mass flux is treated as entirely diffusive, that is, $L(x)=0$. For the friction operator, we choose the local diffusion operator
\begin{equation}\label{Mexpr}
    M = - \partialr \cdot T_0 \, \bm{\Xi} \cdot \partialr ,
\end{equation}
which, for a positive semidefinite symmetric tensor $\bm{\Xi }$, is also positive semidefinite (as can be shown after an integration by parts and after ignoring boundary terms for the purpose of deriving bulk evolution equations). These properties are even more obvious in the bracket notation
\begin{equation}\label{dbrackdiff}
    [A,B] = \int_\Lambda \left( \partialr \frac{\delta A}{\delta\rho} \right)
    \cdot T_0 \, \bm{\Xi} \cdot
    \left( \partialr \frac{\delta B}{\delta\rho} \right) d^3 r .
\end{equation}
The degeneracy requirement (\ref{MEconsistency}) for this dissipative bracket is satisfied because $\delta E/\delta\rho$ is constant for our hard-sphere system. We have now specified all thermodynamic building blocks of the GENERIC framework and are ready to write out the evolution of our diffusive system. From the fundamental equation (\ref{GENERICLM}) we obtain the evolution equation
\begin{equation}\label{diffeq1}
    \frac{\partial\rho}{\partial t} = \partialr \cdot \bm{\Xi}
    \cdot \frac{\partial\mu}{\partial\bm{r}} .
\end{equation}
The gradient of the chemical potential is the natural driving force for diffusive mass transport in an isothermal system. With the help of the explicit form (\ref{entropydens}) of the entropy the driving force can alternatively be formulated in terms of the density gradient,
\begin{equation}\label{diffeq2}
    \frac{\partial\rho}{\partial t} = \partialr \cdot
    \frac{\kB T_0}{m} \, \frac{R_2(\rho)}{\rho} \,
    \bm{\Xi} \cdot \frac{\partial\rho}{\partial\bm{r}}
    = \partialr \cdot \bm{D}(\rho)
    \cdot \frac{\partial\rho}{\partial\bm{r}} ,
\end{equation}
where the function $R_2(\rho)$ results from first and second order derivatives of $R_0(\rho)$,
\begin{equation}\label{R2def}
    R_2(\rho) = 1 -  \frac{d}{d\rho} \left[
    \frac{\rho^2}{R_0(\rho)} \frac{d R_0(\rho)}{d\rho} \right] ,
\end{equation}
and the second part of Eq.~(\ref{diffeq2}) is merely a definition of the diffusion tensor
\begin{equation}\label{Ddef}
    \bm{D}(\rho) =
    \frac{\kB T_0}{m} \, \frac{R_2(\rho)}{\rho} \,
    \bm{\Xi} .
\end{equation}
Even for a constant tensor $\bm{\Xi}$, a complicated dependence of the diffusion tensor on $\rho$ arises. For the ideal gas, we find $R_2=1$ and $\bm{D} \propto 1/\rho$.

The rate of change of the total energy in the system is given by the chain rule \cite{hco162},
\begin{equation}\label{Echange}
    \frac{d E}{d t} = \frac{\delta E}{\delta x} M \frac{\delta S}{\delta x}
    = [E,S] + \frac{3}{2} \frac{\kB T_0}{m} \int_{\partial\Lambda} \bm{n}
    \cdot \bm{D} \cdot \frac{\partial\rho}{\partial\bm{r}} \, d^2r ,
\end{equation}
where $\bm{n}$ is the normal vector on the boundary $\partial\Lambda$ of the domain $\Lambda$ pointing from the inside to the outside of the system. The boundary term is a result of the integration by parts required to go from Eq.~(\ref{Mexpr}) to Eq.~(\ref{dbrackdiff}). As the dissipative bracket $[E,S]$ vanishes according to Eq.~(\ref{MEconsistency}), the energy of our isothermal hard sphere system changes according to the mass flux into the system. For steady state systems, the total mass flux must be zero. We are interested in the nonequilibrium steady state systems arising for a given nonuniform chemical potential on $\partial\Lambda$.

With the bracket in Eq.~(\ref{dbrackdiff}), we have introduced a dissipative mass flux. Such a possibility has been debated controversially in the context of the full Navier-Stokes-Fourier equations of hydrodynamics. In a classical paper, Dzyaloshinskii and Volovick \cite{DzyalVolov80} proposed the inclusion of a dissipative mass flux into the hydrodynamic equations. Starting from a modified kinetic theory, Klimontovich \cite{Klimontovich92} arrived at the same suggestion. More recently, a dissipative contribution to the mass flux was re-introduced and forcefully promoted by Brenner \cite{Brenner06}, whose work stimulated significant interest and controversy in the physics and fluid dynamics communities. Brenner's work motivated thorough investigations on the thermodynamic admissibility of a dissipative contribution to the mass flux \cite{hcobet,hco166,hco164,hco170,hco175}, which demonstrated that the idea fits naturally into the GENERIC framework and into standard linear irreversible thermodynamics. However, these investigations focused entirely on nonequilibrium thermodynamics and neglected other, equally important considerations, such as the local conservation of angular momentum. Such additional criteria were considered earlier in \cite{KostadtLiu98} and corrobated the original (not rigorously justified) statement of Landau and Lifshitz that a dissipative contribution to the mass flux cannot exist (see footnote at the end of Sec.~49 of \cite{LandauLifshitz6}). A concise summary of the current state of the discussion can be found in the comment \cite{Liu08} to the letter \cite{hco175}, in the reply \cite{hco181} to that comment, and in a recent joint manuscript by the same authors \cite{hco161}. Fluctuations in the mass density arise entirely from fluctuations in the momentum and heat fluxes, so that the equations considered in this paper are not of a truly hydrodynamic origin.

\section{Macroscopic Fluctuation Theory}\label{secMFT}
Before turning to macroscopic fluctuation theory, let us summarize some key elements of GENERIC. In the GENERIC framework, the building blocks of nonequilibrium thermodynamics are formulated to construct time-evolution equations for thermodynamic systems. In particular, \emph{the hallmark of nonequilibrium entropy is to generate irreversible dynamics}. For complex fluids, this nonequilibrium entropy depends on additional structural variables, such as polymer conformation tensors in polymeric liquids \cite{hcobet}. In macroscopic fluctuation theory \cite{Bertinietal09} (see also \cite{Bertinietal02,Bertinietal04}), the hydrodynamic equations for averages obtained as an \emph{output} from GENERIC are assumed to be given as an \emph{input} to the theory of nonequilibrium thermodynamics because the interest is in the fluctuations around the solutions of the hydrodynamic equations. For the diffusive system considered in the previous section, it is assumed that one knows the continuity equation for the density
\begin{equation}\label{conteq}
    \frac{\partial\rho}{\partial t} = - \partialr \cdot \bm{j} ,
\end{equation}
with a mass flux $\bm{j} = \bm{j}(\rho)$ given by the constitutive equation
\begin{equation}\label{jMFT}
    \bm{j} = - \bm{D}(\rho) \cdot \frac{\partial\rho}{\partial\bm{r}}
    + \bm{\chi}(\rho) \cdot \bm{E} .
\end{equation}
These two equations correspond to Eq.~(\ref{diffeq2}) with an additional flux contribution resulting from an external field $\bm{E}$, where $\bm{\chi}(\rho)$ is the mobility tensor. The external field can be used to control the evolution of the density profile. The hydrodynamic equations (\ref{conteq}) and (\ref{jMFT}) are supplemented by boundary conditions, more precisely, by specifying the chemical potential $\mu$ on the boundary $\partial\Lambda$ of the domain $\Lambda$.

In macroscopic fluctuation theory, \emph{the hallmark of nonequilibrium entropy is to govern fluctuations}. The goal of this approach is to introduce such an entropy entirely in terms of macroscopic concepts and considerations, with a validation of the phenomenological ideas by microscopic models \cite{Bertinietal06,Bertinietal09}.

In intuitive terms, the fluctuations around stationary states can be understood in the following way (all the details can be found in \cite{Bertinietal02,Bertinietal04,Bertinietal06,Bertinietal09}): Fluctuations decay according to the macroscopic hydrodynamic equations, and they spontaneously emerge according to the time-reversed trajectories governed by ``adjoint hydrodynamics.'' Note that the fluctuations are assumed to satisfy the non-fluctuating boundary conditions and they hence depend on the nature of the boundary conditions. Consideration of stationary states is crucial for the analysis of emerging and decaying fluctuations. In other words, the construction of adjoint hydrodynamics depends on the probability density for fluctuations in the steady state. Only for equilibrium states, adjoint hydrodynamics coincides with hydrodynamics; we then have time-reversal symmetry. The macroscopic time-reversal behavior of diffusion equations has been studied in great detail within the theory of stochastic differential equations in the classical work of Nelson \cite{Nelson} and, more generally, is part of the theory of Markov processes. These time-reversed trajectories have been shown to minimize a suitable cost function for fluctuations, which is naturally identified with the time integral of the extra dissipation rate caused by fluctuations. The trajectories of adjoint or time-reversed hydrodynamics minimize the work required for a dynamic transition from a stationary initial state to a given final state, which can be achieved by suitably chosen external forces. As a result of such a variational principle, a nonequilibrium entropy or free energy associated with fluctuations can be introduced.

In terms of equations, adjoint hydrodynamics for the diffusive system is governed by the continuity equation (\ref{conteq}) with the ``reversed'' mass flux
\begin{equation}\label{jMFTadj}
    \bm{j}^* = \bm{D}(\rho) \cdot \frac{\partial\rho}{\partial\bm{r}}
    - \bm{\chi}(\rho) \cdot \left[ \bm{E} + 2 \partialr
    \frac{\delta {\cal F}(\rho,\bar{\rho})}{\delta\rho} \right] ,
\end{equation}
instead of the expression in Eq.~(\ref{jMFT}). The quantity ${\cal F}(\rho,\bar{\rho})$ introduced in Eq.~(\ref{jMFTadj}) is referred to as the nonequilibrium free energy of the macroscopic state $\rho$ for a system in the stationary state $\bar{\rho}$  \cite{Bertinietal09}. It describes the probability distribution for fluctuations $p_{\bar{\rho}}(\rho)$ around the stationary state $\bar{\rho}$,
\begin{equation}\label{Fproprel}
    \frac{p_{\bar{\rho}}(\rho)}{p_{\bar{\rho}}(\bar{\rho})} =
    \exp \left\{ - \frac{{\cal F}(\rho,\bar{\rho})}{\kB T_0} \right\} .
\end{equation}
The occurrence of a functional ${\cal F}(\rho,\bar{\rho})$ in Eqs.~(\ref{jMFTadj}) and (\ref{Fproprel}) allows for the possibility of nonlocal time reversal and fluctuation effects.

For the actual calculation of ${\cal F}(\rho,\bar{\rho})$, one considers the extra dissipation rate as a Lagrangian and passes by Legendre transformation to the Hamiltonian formulation. The associated Hamilton-Jacobi equation is then given by \cite{Bertinietal02,Bertinietal04,Bertinietal09}
\begin{equation}\label{HamJaceq}
    \int_\Lambda \left[ \left( \partialr \frac{\delta{\cal F}}{\delta\rho} \right) \cdot \bm{\chi}(\rho) \cdot \left( \partialr \frac{\delta{\cal F}}{\delta\rho} \right) - \frac{\delta{\cal F}}{\delta\rho} \partialr \cdot \bm{j}(\rho) \right] d^3r = 0 .
\end{equation}
The Hamilton-Jacobi equation (\ref{HamJaceq}) is a convenient starting point for the practical calculation of ${\cal F}(\rho,\bar{\rho})$, be it by verification of guessed solutions or by perturbation theory. It can be rewritten as
\begin{equation}\label{HamJaceqdi}
    \int_\Lambda \frac{\delta{\cal F}}{\delta\rho} \left[ \partialr \cdot
    \bm{\chi}(\rho) \cdot \partialr \frac{\delta{\cal F}}{\delta\rho}
    + \partialr \cdot \bm{j}(\rho) \right] d^3r =
    \int_{\partial\Lambda} \frac{\delta{\cal F}}{\delta\rho} \, \bm{n}
    \cdot \bm{\chi}(\rho) \cdot \partialr \frac{\delta{\cal F}}{\delta\rho} \, d^2r ,
\end{equation}
which is particularly convenient when $\delta{\cal F}/\delta\rho$ vanishes on the boundary so that the surface integral on the right-hand side is equal to zero.

In general, the solution ${\cal F}(\rho,\bar{\rho})$ depends on the dynamic material properties $\bm{\chi}(\rho)$ and $\bm{D}(\rho)$ occurring in the variational problem. In blatant contrast, the statistical expressions for the energy $E$, the entropy $S$, and hence also for the free energy $F$ of GENERIC (as obtained by means of projection operator techniques) imply that all these quantities can be expressed in terms of a nonequilibrium ensemble and hence do not contain any dynamic material information.

For homogeneous or, in the presence of external fields, inhomogeneous equilibrium states $\bar{\rho}$ characterized by $\bm{j}(\bar{\rho}) = 0$, one finds \cite{Bertinietal09}
\begin{equation}\label{Feq}
    {\cal F}(\rho,\bar{\rho}) = \int_\Lambda \left[ f(\rho(\bm{r})) - \bar{\mu}(\bm{r}) \rho(\bm{r}) \right] d^3r
    - \int_\Lambda \left[ f(\bar{\rho}(\bm{r})) - \bar{\mu}(\bm{r}) \bar{\rho}(\bm{r}) \right] d^3r .
\end{equation}
The occurrence of the Legendre transform of the Helmholtz free energy density $f$ is natural because the problem is controlled by the chemical potential on the boundaries. Note that, in the first integral in Eq.~(\ref{Feq}), there actually occurs the chemical potential $\bar{\mu}(\bm{r})$ associated with the background stationary state $\bar{\rho}(\bm{r})$ in front of $\rho(\bm{r})$. This causes a nontrivially coupled dependence of ${\cal F}$ on $\rho$ and $\bar{\rho}$ and is important to obtain a functional derivative that vanishes on the boundary,
\begin{equation}\label{Feqdr}
    \frac{\delta{\cal F}}{\delta\rho} = \mu(\bm{r}) - \bar{\mu}(\bm{r}) ,
\end{equation}
which is needed in the verification of the Hamilton-Jacobi equation (\ref{HamJaceqdi}), together with the Nernst-Einstein relation
\begin{equation}\label{NernstEinst1}
    \bm{\chi}(\rho) \, \frac{\partial^2 f}{\partial \rho^2} = \bm{D}(\rho) .
\end{equation}
Note that $\rho \, \partial^2 f/\partial \rho^2$ is the isothermal speed of sound squared and hence positive. In terms of the more natural diffusion tensor $\bm{\Xi}$ introduced via the friction matrix (\ref{Mexpr}) of the GENERIC approach, the Nernst-Einstein relation takes the simpler form
\begin{equation}\label{NernstEinst2}
    \bm{\chi}(\rho) = \bm{\Xi}(\rho) .
\end{equation}
The condition $\bm{j}(\bar{\rho}) = 0$ for equilibrium is equivalent to the time-reversal symmetry $\bm{j}(\rho) = \bm{j}^*(\rho)$ \cite{Bertinietal09}.

The Legendre transform of ${\cal F}(\rho,\bar{\rho})$ from the density $\rho$ to its conjugate variable, the chemical potential $\mu$, provides the generating functional for correlation functions. This approach has been elaborated in Section~4 of \cite{Bertinietal09}.

\section{Thermodynamics: Averages and Fluctuations}\label{secavefluc}
As we have emphasized in the preceding sections, GENERIC deals with the evolution equations for macroscopic averages, whereas macroscopic fluctuation theory deals with the fluctuations around steady state averages. GENERIC is of little interest in formulating hydrodynamic equations because the structure of these equations is fixed by the local conservation laws for mass, momentum and energy. The more ambitious goal of GENERIC is to develop proper and consistent equations for complex fluids with additional slow structural variables, for which there are no conservation laws; then, GENERIC provides helpful structural guidance. For complex fluids, the equations for the averages need to be formulated first, and a theory of fluctuations can only be developed as a second step. Only for conserved quantities, the balance equations for the averages are so straightforward that one can proceed directly to a theory of macroscopic fluctuations. We hence refer to \emph{nonequilibrium thermodynamics of the first and second kind}. From a statistical mechanics perspective, one could also say that thermodynamics of the first kind is associated with the law of large numbers for averages, whereas thermodynamics of the second kind is associated with the central limit theorem \cite{Bertinietal06}.

There are a number of further associations that should go with the distinction of thermodynamics of the first and second kind that go beyond the emphasis of a natural sequence of steps. In the case of small Gaussian fluctuations, the averages of thermodynamics of the first kind are given by the first moments, whereas the fluctuations of thermodynamics of the second kind are described by the second moments. There is also a correspondence to the fluctuation dissipation relations of the first and second kind distinguished by Kubo \cite{KuboetalII}. The fluctuation dissipation relation of the first kind deals with the average response of a system to an external perturbation and belongs to the field of linear response theory. The fluctuation dissipation relation of the second kind provides information about the second moments of the noise in a stochastic description of a system and is deeply linked to the projection operator approach.

At equilibrium, of course, thermodynamics of the first and second kind are related by Einstein's theory of fluctuations (see, for example, Section 10.B of \cite{Reichl} or, more generally, Chapter~19 of \cite{Callen}). Far away from equilibrium, an equivalence of thermodynamics of the first and second kind is far from obvious. A deeper comparison between GENERIC and the macroscopic fluctuation theory based on fluctuation dissipation relations is attempted in the subsequent section. Among the other attempts to establish a general theory of nonequilibrium steady states mentioned in the introduction, the work of Oono and Paniconi (1998) \cite{OonoPaniconi98} and of Sasa and Tasaki (2006) \cite{SasaTasaki06} may be classified as thermodynamics of the first kind, whereas the work of Taniguchi and Cohen (2007) \cite{TaniguchiCohen07} deals with thermodynamics of the second kind.

The distinction elaborated in the present section is also useful for the discussion of different \emph{variational principles}. On the one hand, the so-called principle of minimal entropy production of linear irreversible thermodynamics \cite{Prigogine,deGrootMazur,Kreuzer}, that is, within thermodynamics of the first kind, is of limited validity (see, for example, p.~832 of \cite{Bertinietal04} or Section 3.1.5 of \cite{hcobet}). On the other hand, the minimum dissipation principle assumed in the construction of macroscopic fluctuation theory, which was briefly discussed in the paragraph before Eq.~(\ref{jMFTadj}), seems to be of more general validity.

\section{Fluctuation Dissipation Relations}\label{secFDR}
Nonequilibrium thermodynamics of the first and second kind focus on averages and fluctuations, respectively. In the GENERIC formulation of the equations for averages, an important feature is the proper formulation of irreversible or dissipative dynamics (after separating it from reversible dynamics). It is hence natural to look at fluctuation dissipation relations in order to establish a connection between GENERIC and macroscopic fluctuation theory. Each of the approaches has offered such relations.

\subsection{From Macroscopic Fluctuation Theory to GENERIC}
In the framework of macroscopic fluctuation theory, there exists a so-called \emph{nonlinear fluctuation dissipation relation} for stationary nonequilibrium states. From the equations of Section~\ref{secMFT}, a version of this nonlinear fluctuation dissipation relation is obtained as the sum of Eqs.~(\ref{jMFT}) and (\ref{jMFTadj}),
\begin{equation}\label{nonlinFDR1}
    \frac{1}{2} (\bm{j} + \bm{j}^*) = - \bm{\chi}(\rho) \cdot \partialr
    \frac{\delta {\cal F}}{\delta\rho} .
\end{equation}
This equation (see also Eq.~(2.15) of \cite{Bertinietal02}) is reminiscent of the irreversible contribution to GENERIC in Eq.~(\ref{GENERICLM1}), however, it is the nonequilibrium free energy  governing fluctuations that generates the arithmetic mean of real hydrodynamic and adjoint hydrodynamic evolution characterizing the relaxation and emergence of fluctuations around a given stationary state, respectively (for equilibrium states with $\bm{j} = \bm{j}^*$, the entropies of GENERIC and of macroscopic fluctuation theory  hence coincide). The reformulation of the nonlinear fluctuation dissipation relation to obtain an equation for the relaxation of fluctuations in the top half of p.~645 of \cite{Bertinietal02} is even more similar to GENERIC,
\begin{equation}\label{nonlinFDR2}
    \frac{\partial\rho}{\partial t} = - \partialr \cdot \bm{j} =
    {\cal A}(\rho) + M \frac{\delta {\cal S}}{\delta\rho} ,
\end{equation}
where $\cal S$ is the entropy of macroscopic fluctuation theory and ${\cal A}(\rho)$ is a vector field that conserves the entropy,
\begin{equation}\label{entropcons}
    \int_\Lambda \frac{\delta{\cal S}}{\delta\rho} \, {\cal A}(\rho) \, d^3r = 0 .
\end{equation}
Equation (\ref{entropcons}) corresponds to the degeneracy (\ref{LSconsistency}) of the GENERIC framework, which is a hallmark of reversibility of the vector field ${\cal A}(\rho)$. GENERIC additionally postulates a degenerate Poisson structure to formulate the entropy-conserving or reversible vector field ${\cal A}(\rho)$ in terms of a Hamiltonian. Once more we emphasize that, in spite of striking similarities between Eqs.~(\ref{GENERICLM}) and (\ref{nonlinFDR2}), the entropy  $\cal S$ of the macroscopic fluctuation theory depends on the underlying steady state and on dynamic material properties, whereas the entropy $S$ of GENERIC does not.

If we neglect the surface term in the Hamilton-Jacobi equation (\ref{HamJaceqdi}) and rewrite it in the compact form
\begin{equation}\label{HamJaceqx}
    \int_\Lambda \frac{\delta{\cal F}}{\delta\rho} \left[ - \partialr
    \cdot \bm{j}(\rho) + \frac{1}{T_0} M
    \frac{\delta {\cal F}}{\delta\rho} \right] d^3r = 0 ,
\end{equation}
then we arrive at an alternative interpretation of this equation inspired by the GENERIC framework. The free energy $\cal F$ is to be constructed such that it generates irreversible dynamics and, at the same time, is conserved under the reversible dynamics remaining after subtraction of the irreversible contribution from the full dynamic equation. This double role of the free energy is the reason for the quadratic occurrence of $\delta{\cal F}/\delta\rho$ in Eq.~(\ref{HamJaceqx}). Also note that the dissipative bracket $[{\cal F},{\cal F}]$ of Eq.~(\ref{dbrackdiff}) occurs in the full Hamilton-Jacobi equation (\ref{HamJaceq}).

According to the GENERIC approach, when formed with a constant temperature $T_0$, the free energy $F(\rho)$ actually generates irreversible dynamics and is conserved so that ${\cal F} = F$ satisfies Eq.~(\ref{HamJaceqx}). However, $\delta F/\delta\rho$ does not vanish on the boundary, as was assumed in order to arrive at Eq.~(\ref{HamJaceqx}). The difference between the nonequilibrium free energies of GENERIC and the macroscopic fluctuation theory hence results exclusively from the boundary condition $\delta F/\delta\rho = 0$. The complicated functional dependence of $\cal F$ on the underlying steady state $\bar{\rho}$ and on dynamic material properties is recognized as entirely due to the boundary conditions propagating into our driven diffusive system to achieve steady state conditions. In short, the fluctuations of the steady state system are complicated because no fluctuations of the chemical potential on the boundary are allowed.

\subsection{From GENERIC to Macroscopic Fluctuation Theory}
We have seen that the macroscopic fluctuation theory comes with GENERIC-type equations, but with a more complicated entropy resulting from boundary effects. We can now ask the reverse question: Can GENERIC provide equations for fluctuations around the solutions of the equations for averages? Fortunately, GENERIC is backed up by statistical mechanics and a corresponding fluctuation dissipation relation.

The statistical approach to GENERIC is based on projection operator techniques and nonequilibrium ensembles (see Section 6.1.2 of \cite{hcobet}), and the choice of the ensemble clearly corresponds to an \emph{assumption} about the nonequilibrium fluctuations. This may be considered as a weakness of the statistical approach to GENERIC. An analogous situation arises in Einstein's theory of equilibrium fluctuations where one needs to argue that the extensive quantities fluctuate for given intensive variables, which are fixed by a surrounding bath. Landau and Lifshitz are willing to assume an isolated system at constant energy to justify a fluctuating temperature \cite{LandauLifshitz5}. For a generalized canonical nonequilibrium ensemble, the probability distribution for fluctuations $p_{\bar{\rho}}(\rho)$ around a given state $\bar{\rho}$ occurring in Eq.~(\ref{Fproprel}) has actually been discussed in a more general context in Exercise 138 of \cite{hcobet}. The resulting nonequilibrium free energy associated with the fluctuations of our diffusive system can be written as
\begin{equation}\label{FeqG}
    {\cal F}(\rho,\bar{\rho}) = F(\rho) - F(\bar{\rho}) - \int_\Lambda
    \bar{\mu}(\bm{r}) \left[ \rho(\bm{r}) - \bar{\rho}(\bm{r}) \right] d^3r ,
\end{equation}
which is a direct generalization of the equilibrium free energy in Eq.~(\ref{Feq}). Such a simple local equilibrium generalization, and hence the concept of generalized canonical nonequilibrium ensembles, would be inappropriate according to the macroscopic fluctuation theory. The deeper reason is that intensive variables can be prescribed only on the boundaries and not throughout the bulk system. Note, however, that macroscopic fluctuation theory in its usual form relies on non-fluctuating boundary conditions for the chemical potential which may be challenged, at least from a phenomenological perspective (and one may try to justify them by statistical mechanics). The intensive quantities are related to the Lagrange multipliers of the canonical ensemble; as such, they are a property of the entire ensemble rather than individual fluctuations. Even at equilibrium, the naturally fluctuating variables are the extensive quantities, but fluctuations of intensive variables like temperature or chemical potential are often introduced through their thermodynamic equations of state (see, for example, Eq.~(10.14) of \cite{Reichl} or \S~112 of \cite{LandauLifshitz5}; on the other hand, in Chapter~19 of his textbook \cite{Callen}, Callen refrains from introducing fluctuations of intensive variables).

A more fundamental approach to the fluctuations to be added to GENERIC is obtained by using probability densities of the original variables as new variables in the projection operator approach (see Section 6.3 of \cite{hcobet}). It is found that the generalized canonical ensemble on the level of probability densities is intimately related to the generalized microcanonical ensemble on the level of the original variables. As a result of the projection operator procedure, one obtains the very natural and appealing Fokker-Planck equation,
\begin{eqnarray}
  \frac{\partial p(x,t)}{ \partial t} &=&
  - \frac{\delta}{\delta x} \left[ \left(
  L(x) \frac{\delta E(x)}{\delta x} +
  M(x) \frac{\delta S(x)}{\delta x} \right) p(x,t) \right] \nonumber\\
  &+& \kB \frac{\delta}{\delta x} \left[ M(x)
  \frac{\delta}{\delta x} p(x,t) \right] ,
\label{FPEGEN}
\end{eqnarray}
governing the dynamics of fluctuations in arbitrary nonequilibrium systems, where all GENERIC building blocks must be evaluated in the fundamental microcanonical ensemble. Equation (\ref{FPEGEN}) consists of a GENERIC drift and a superimposed multiplicative white noise \cite{Gardiner,hcobook}. The representation of the solutions of Fokker-Planck equations in terms of functional integrals and the development of field-theoretic solution methods can be found in the textbook \cite{Honerkamp}; such a representation is particularly useful for the interpretation of Eq.~(\ref{FPEGEN}) in terms of fluctuating trajectories.

For the evolution of the average of an arbitrary observable $A$, the Fokker-Planck equation (\ref{FPEGEN}) implies
\begin{equation} \label{brackformave}
  \frac{d\aver{A}}{dt} = \aver{\{A,E\}} + \aver{[A,S]}
  + \kB \aver{\frac{\delta}{\delta x} M \frac{\delta A}{\delta x}} .
\end{equation}
This equation can be used to evaluate averages and correlations in time-dependent and steady state situations.

We are particularly interested in fluctuations with experimentally observable consequences, most importantly in scattering experiments \cite{OrtizSengers}. Except near critical points, fluctuations are usually very small which, of course, is important for the success of thermodynamics. At equilibrium, Einstein's famous fluctuation theory hence relies on a second-order expansion of the entropy leading to Gaussian fluctuations. Whereas large fluctuations can be treated easily and elegantly (see Chapter~19 of \cite{Callen}), there is usually no need to go beyond Gaussian fluctuations. In the same spirit, the Fokker-Planck equation (\ref{FPEGEN}) of GENERIC with fluctuations can handle all problems involving small nonequilibrium fluctuations. A special case of the Fokker-Planck equation (\ref{FPEGEN}) corresponds to fluctuating hydrodynamics. Whereas the correlations of the non-conserved fluxes are short range, in nonequilibrium systems, the well-known long range fluctuations of conserved quantities arise even after linearization of the noise terms \cite{OrtizSengers}.

If we are interested in small fluctuations around steady states, we make the simplifying assumption that the averages $\bar{x}$ are characterized by the condition of vanishing macroscopic time evolution,
\begin{equation}\label{ssdef}
  \frac{d\bar{x}}{dt} = L(\bar{x}) \frac{\delta E(\bar{x})}{\delta \bar{x}} +
  M(\bar{x}) \frac{\delta S(\bar{x})}{\delta \bar{x}} = 0 .
\end{equation}
In other words, we do not take fluctuation renormalization \cite{Grabert,hcobet} into account. It is then natural to assume additive or linearized noise, that is, to replace $M(x)$ by $M(\bar{x})$ in the last term of Eqs.~(\ref{FPEGEN}) and (\ref{brackformave}). From Eq.~(\ref{brackformave}) we then obtain a powerful equation for calculating steady state correlations of small fluctuations of $x$ around $\bar{x}$,
\begin{equation}\label{powerss}
    \averr{x_i,\dot{x}_j} + \averr{\dot{x}_i,x_j} + 2 \kB M_{ij}(\bar{x}) = 0 ,
\end{equation}
where $\averr{\cdot,\cdot}$ denotes the covariance of two observables and $\dot{x}$ is the linearized right-hand side of the GENERIC evolution equation (\ref{GENERICLM}). The linearized GENERIC evolution involves reversible dynamics and the entropy in addition to the friction matrix $M$. For the diffusion problem, it is the discrepancy between $\bm{\Xi}$ in the friction matrix in Eq.~(\ref{Mexpr}) and $\bm{D}$ in the evolution equation (\ref{diffeq2}) that produces interesting correlations. According to Eq.~(\ref{NernstEinst1}), this discrepancy is determined by the behavior of the local equilibrium free energy. In hydrodynamics, the reversible convection effects occurring in the first two terms of Eq.~(\ref{powerss}) lead to long range density correlations \cite{OrtizSengers}.

As mentioned at the end of Section \ref{secGENERIC}, the assumption of isothermal conditions may be subtle and might require a more careful discussion because, except in proper limiting cases, a constant $T_0$ is inconsistent with the steady state conditions (\ref{ssdef}) and (\ref{powerss}). We tried to circumvent this problem by considering athermal hard-sphere systems for which a GENERIC formulation without a temperature variable works well. However, a dependable discussion of the local entropy production and flux is clearly desirable.

\subsection{One-Dimensional Simple Exclusion Process}
In order to illustrate the calculation of long range nonequilibrium correlations within the GENERIC framework with fluctuations, we consider one-dimensional diffusion in the interval $[0,l]$. We assume a constant diffusion coefficient $D$, so that the steady state solutions of the diffusion equation (\ref{diffeq2}) are of the form
\begin{equation}\label{steady1d}
    \bar{\rho}(r) = \rho_0 + (\rho_l - \rho_0) \frac{r}{l} ,
\end{equation}
where $\rho_0$ and $\rho_l$ are the values of the density profile $\bar{\rho}$ at the boundaries of the interval $[0,l]$. For the quantity $\Xi$ we assume a particular form of Eq.~(\ref{Ddef}),
\begin{equation}\label{Ddef1d}
    \Xi = \rho \left( 1 - \frac{\rho}{\rho_{\rm c}} \right) \frac{m}{\kB T_0} \, D .
\end{equation}
If the parameter $\rho_{\rm c}$ in this equation goes to infinity, we recover ideal gas behavior. A finite value of $\rho_{\rm c}$ corresponds to the simple exclusion process of \cite{Bertinietal06}. Note, however, that all our calculations are based on the phenomenological equation (\ref{ssdef}) and the fluctuation dissipation relation (\ref{powerss}).

If we apply Eq.~(\ref{powerss}) to $x_i = \rho(r)$, $x_j = \rho(s)$ and introduce the covariance function
\begin{equation}\label{covdef}
    {\rm cov}(r,s) = \averr{\rho(r),\rho(s)} ,
\end{equation}
then we obtain the equation
\begin{equation}\label{coveq}
    \left( \frac{\partial^2}{\partial r^2}
    + \frac{\partial^2}{\partial s^2} \right) {\rm cov}(r,s) =
    2 m \frac{\partial}{\partial r} \bar{\rho}(r)
    \left( 1 - \frac{\bar{\rho}(r)}{\rho_{\rm c}} \right)
    \frac{\partial}{\partial r} \delta(r-s) .
\end{equation}
The most general solution of this equation is given by
\begin{eqnarray}
    {\rm cov}(r,s) &=& m  \bar{\rho}(r) \left( 1 - \frac{\bar{\rho}(r)}{\rho_{\rm c}} \right)
    \delta(r-s) \nonumber\\
    &-& \frac{m (\rho_l - \rho_0)^2}{\rho_{\rm c} l^2}
    \left[ \frac{r+s}{2} - \frac{|r-s|}{2} - \frac{rs}{l} \right] + \phi(r,s) ,
\label{covsol1}
\end{eqnarray}
where $\phi(r,s)$ is a solution of the two-dimensional Laplace equation on the square $[0,l]\times[0,l]$ determined by the desired boundary conditions. We hence actually find the previously mentioned freedom of choosing the boundary conditions. Nonfluctuating boundary conditions imply $\phi(r,s)=0$. For $r>s$, we then have the equation
\begin{equation}\label{cosol2}
    {\rm cov}(r,s) = - (\rho_l - \rho_0)^2 \frac{m}{l \rho_{\rm c}}
    \, \frac{s}{l} \left( 1 - \frac{r}{l} \right) ,
\end{equation}
which expresses the negative nonlocal correlations for the simple exclusion process given in Eq.~(2.3) of \cite{Bertinietal06}. Note that these nonlocal correlations vanish in the limit $\rho_{\rm c} \rightarrow \infty$, that is, for the ideal gas. The exclusion mechanism is crucial for obtaining long range fluctuations in nonequilibrium steady states, that is, for $\rho_l \neq \rho_0$. In the GENERIC approach, there is no need to introduce a nonlocal entropy functional to obtain the correlations (\ref{covsol1}) form purely phenomenological equations. The GENERIC nonequilibrium entropy is of the local equilibrium form. The origin of this simplicity lies in the fact that the relationship (\ref{NernstEinst1}) depends only on local equilibrium thermodynamic properties.

We once more emphasize the possibility of choosing different boundary conditions for $\phi(r,s)$. This implies different correlations and hence different nonequilibrium free energies according to macroscopic fluctuation theory.

\section{Summary and Discussion}\label{secsumdisc}
In the context of an isothermal driven diffusive system, we have compared two different approaches to nonequilibrium systems, one based on averages and the other one on fluctuations. To emphasize this conceptual difference between the approaches, we have classified the GENERIC framework, which is focused on the structure of time-evolution equations for averages, and the macroscopic fluctuation theory as thermodynamics of the first and second kind, respectively. In general, one first needs to find the proper equations for the averages before, in a second step, one can discuss the fluctuations around them. Deep relationships between the two approaches can be revealed by means of the fluctuation dissipation relations that exist in either approach.

Both approaches introduce nonequilibrium entropies and Helmholtz free energies. Whereas the free energy of GENERIC contains only static material information, the free energy of the macroscopic fluctuation theory depends on transport coefficients. This difference is revealed to be a consequence of boundary conditions which, in macroscopic fluctuation theory, are assumed to be non-fluctuating. Different boundary conditions lead to different free energies.

The hydrodynamic equation for the decay of fluctuations in the macroscopic fluctuation theory formally possesses the GENERIC structure, and the Fokker-Planck equation governing fluctuations within the statistically founded GENERIC approach provides the counterpart to the Hamilton-Jacobi equation for the nonequilibrium free energy of the macroscopic fluctuation theory. The steady state moment equation (\ref{powerss}) resulting from this Fokker-Planck equation provides a powerful tool for calculating long range nonequilibrium correlations. An explicit calculation for a one-dimensional model shows that GENERIC can reproduce the long range nonequilibrium correlations of macroscopic fluctuation theory in spite of the different nonequilibrium entropies used in the two approaches. The origin of these long range correlations for a model with dissipative mass flux, which is not truly of hydrodynamic origin, lies in the local equilibrium thermodynamic properties and their position dependence resulting from a nonuniform density profile.

It is well known that the different nonequilibrium ensembles used in the projection operator derivation of GENERIC may at best be equivalent for the resulting average equations but certainly not for the description of fluctuations. Macroscopic fluctuation theory suggests that the generalized canonical ensemble does not represent fluctuations in a boundary-driven system in an appropriate way. However, this conclusion relies on the debatable assumption of non-fluctuating boundary conditions. A more systematic projection operator based theory of fluctuations, even for time-dependent nonequilibrium systems, is given by the Fokker-Planck equation (\ref{FPEGEN}). A detailed discussion shows the possibility of choosing from a variety of different boundary conditions. An even more illuminating and conclusive comparison between GENERIC and macroscopic fluctuation theory might be obtained for flow-driven systems, where the driving mechanism is felt via a velocity field throughout a bulk system rather than at the boundaries only and a reliable complete set of hydrodynamic equations is used.

This paper is a plea to look for simplicity in beauty and generality rather than in propitious special cases. Eventually, fully consistent results should be achieved.

\begin{acknowledgements}
This work has been stimulated at the workshop on ``Open Systems: Non-Equilibrium Phenomena---Dissipation, Decoherence, Transport'' organized by J\"urg Fr\"ohlich and Gian Michele Graf at the ETH Z\"urich in June 2009. In particular, I am grateful for the encouraging remarks of David Ruelle.
\end{acknowledgements}


\end{document}